# Leveraging Optical Communication Fiber and AI for Distributed Water Pipe Leak Detection

Huan Wu, Huan-Feng Duan, Wallace W. L. Lai, Kun Zhu, Xin Cheng, Hao Yin, Bin Zhou, Chun-Cheung Lai, Chao Lu, and Xiaoli Ding

*Industry Abstract*— Detecting leaks in water networks is a costly challenge. This article introduces a practical solution: the integration of optical network with water networks for efficient leak detection. Our approach uses a fiber-optic cable to measure vibrations, enabling accurate leak identification and localization by an intelligent algorithm. We also propose a method to access leak severity for prioritized repairs. Our solution detects even small leaks with flow rates as low as 0.027 L/s. It offers a cost-effective way to improve leak detection, enhance water management, and increase operational efficiency.

*Abstract*— Water distribution networks (WDNs) are essential infrastructure for providing fresh water to communities, but detecting leaks for WDNs is challenging and costly. In this article, we propose a novel solution that combines an optical network and WDN for distributed water pipe leak detection. Our approach involves using a standard outdoor fiber-optic cable for distributed vibration measurement along a 40-meter water pipe. To accurately identify and locate leaks, we introduce a leak identification algorithm based on 3D-convolutional neural networks (3D-CNNs) that consider the temporal, spectral, and spatial information. Additionally, we propose a leak quantification method that can help prioritize repairs based on the severity of the leak. We evaluate our scheme for different conditions and find that it can detect leak flow rates as low as 0.027 L/s with a location accuracy of within 3 meters and a quantification accuracy of over 85%. Our proposed method offers a cost-effective and value-added solution for designing optical networks and WDNs in new development areas.

## I. INTRODUCTION

PIPE leakage is a plaguing issue in the throes of the water crisis encountered in many countries. Typically, 20-30% of the water in the pipes is leaked [1]. Solving the critical water loss problem is a challenging task due to the following reasons. Firstly, water pipes are geographically spread over vast distances, for example, in Hong Kong over 8,270 km of pressurized freshwater pipes are in service. However, the inspection length of most leak detection technologies is very limited. Secondly, most of the water pipes are buried underground and relatively inaccessible. A leak can go undetected for many years until it is discovered or develops into a significant burst. Lastly, water pipes with different materials, diameters, pressure, flow conditions, leak geometries, and sizes behave uniquely, making it difficult to determine leak signatures.

Leak detection technologies are key to reducing water loss. Table I provides an overview of the current technologies for water pipe leak detection at both industrial and research stages [2]–[4]. In industry, leak locating methods like leak noise correlator, noise logger, and Smart Ball are initially utilized to narrow down the leak area. Subsequently, pin-point technologies such as listening devices, ground penetration radar (GPR), and infrared thermography are followed to determine the exact leak position. Additionally, other technologies such as transient-based technique that exploits hydraulic behavior and time-domain reflectometry (TDR), which measures the reflection coefficient, have been developed. Optical fiber sensors (OFSs) that use fiber-optic cables as sensing elements to detect physical, chemical or biological properties of the surrounding environment are becoming increasingly popular in the water industry. This is due to their ability to perform long-term monitoring, transmit data at high speeds, and operate without the need for a local power supply. OFSs can be broadly classified into two categories: point sensors and distributed sensors. Point OFSs include wavelength demodulation sensors utilizing fiber Bragg gratings (FBGs) and phase-demodulation sensors based on fiber interferometers. These sensors can directly measure strain and temperature or be designed as pressure sensors, accelerometers, hydrophones, or flow meters. They can be multiplexed to form a network with complex interrogation schemes, and their inspection length mainly depends on the interrogation method. Distributed optical fiber sensors (DOFSs), in contrast, can turn a standard optical communication fiber into hundreds even thousands of sensors with a single interrogator. These sensors can be implemented in the frequency domain or time domain. Rayleigh based optical frequency-domain reflectometry (OFDR) is ideal for pin-pointing leaks, as it can measure strain/temperature in millimeter resolution over ten meters. Conversely, interrogators based on optical time-domain reflectometer (OTDR) have much longer sensing capability and are suitable for locating leaks. Raman based distributed temperature sensor (DTS) and Brillouin based distributed temperature/strain sensor (DTSS) can measure temperature along several tens of kilometers water pipe with meter level resolution. However, leaks from freshwater pipe may not introduce obvious temperature anomalies. For pressurized freshwater pipe leak detection, Rayleigh based distributed acoustic sensor (DAS) shows the most significant potential. In this study, we present a novel approach to detect water pipe leaks by combining the optical networks and water distribution network (WDNs). Our proposed method aims to be non-invasive to the WDN's normal operation and user-friendly for water industry practitioners. Different from previous leak detection method based on DAS [5,6], we install the fiber-optic cable on the outside surface of the water pipe to address concerns about water safety and sensing sensitivity. Additionally, we investigate the leak detection capability of the technology by testing different pipe

**Table I. Current Technologies for Water Pipe Leak Detection**

| Stage | Measurand | Technology | Inspection length | Long-term monitoring | Inline or Non-intrusive | Leak level quantification | Power supply[*1] | Hardware Cost (USD)[*2] |
|---|---|---|---|---|---|---|---|---|
| Industry | Vibration/ pressure | Leak noise correlator | Sensor-to-sensor spacing ~200 m | ✗ | Both | ✗ | ✓ | ~10,000 |
| | | Leak noise logger | Sensor-to-sensor spacing ~200 m | ✓ | Both | ✗ | ✓ | ~1,000 |
| | Acoustic noise | Smart Ball | ~20 km | ✗ | Inline | ✗ | ✓ | N.A. |
| | | Listening devices | Pin-point | ✗ | Non-intrusive | ✗ | Both | ~1,000 |
| | GPR image | Ground penetration radar | Pin-point | ✗ | Non-intrusive | Research stage | ✓ | ~20,000 |
| | Thermography | Infrared thermography | Pin-point | ✗ | Non-intrusive | ✗ | ✓ | ~500 |
| Research | Pressure | Transient-based technique | ~2 km | ✗ | Non-intrusive | Research stage | ✓ | ~10,000 |
| | Reflection coefficient | Time-domain reflectometry | Single sensing element ~100 m | ✓ | Non-intrusive | ✗ | ✓ | ~3,000 |
| | Strain/ Temperature | FBG | N.A. | ✓ | Both | Research stage | ✗ | ~30,000 |
| | | Optical fiber interferometer-based sensor | N.A. | ✓ | Both | Research stage | ✗ | ~30,000 |
| | | Rayleigh based OFDR sensor | ~10 m | ✓ | Both | Research stage | ✗ | ~100,000 |
| | | Brillouin based DTSS | ~50 km | ✓ | Non-intrusive | ✗ | ✗ | ~300,000 |
| | Temperature | Raman based DTS | ~10 km | ✓ | Non-intrusive | ✗ | ✗ | ~50,000 |
| | Vibration | Rayleigh based DAS | ~30 km | ✓ | Both | N.A. | ✗ | ~200,000 |

[*1] It means power supply near the sensor.
[*2] Manpower expenditures are not included. For optical fiber sensors, the cost only includes the interrogator.

flow rates and leak flow rates. To enhance the leak detection's intelligence and autonomy, we leverage deep learning techniques to process the densely distributed time-position vibration signals collected from DAS. Our study seeks to contribute to the development of efficient and effective leak detection methods for WDNs and provide a new way of designing optical networks in a cost-effective and value-added manner.

The integration of optical networks and WDNs comprises four layers, as depicted in Figure. 1. The perception layer involves the use of optical communication fiber to sense the water pipe. It must be installed in a way that captures vibrations induced by leaks from pipes that are either buried underground or installed in utility tunnels. The network layer consists of both the optical network and WDN. These point-to-multipoint provide interconnection services or freshwater to customers' premises. Passive optical networks (PONs) are commonly used to bring optical fiber cabling from the central office (CO) to the end users, with a transmission distance limit of 20 km but can be extended to over 50 km with the Super-PON [7]. If dark fiber is available, it can be used for sensing. Otherwise, communication and sensing can function independently in the same fiber, without interference, based on wavelength division multiplexing technology [8]. The WDN is divided into district metered areas (DMAs) which are between 500 to 3000 properties to manage the pressure and ensure the reliability of water supply reliability. The signal processing layer involves the use of supervisory control and data acquisition (SCADA) systems, commonly used in WDN for gathering, aggregating, and processing data from pressure, flow rate sensors, and controllers. The sensing data from the optical network should be integrated into the SCADA system. Finally, the application layer involves specific applications for the water industry, such as a geographical information system (GIS), to determine whether remedial actions are needed. The design and implementation of these four layers pose various challenges at different stages and will involve private companies and government departments. This article focuses on the feasibility and performance of the technical aspects of the proposed scheme, mainly in the perception and data processing layers.

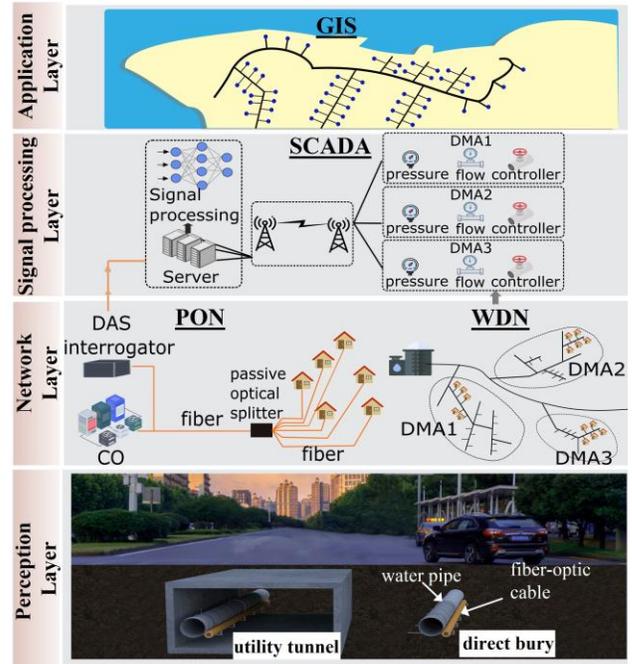

Figure 1. Architecture of smart water leak detection WDN.

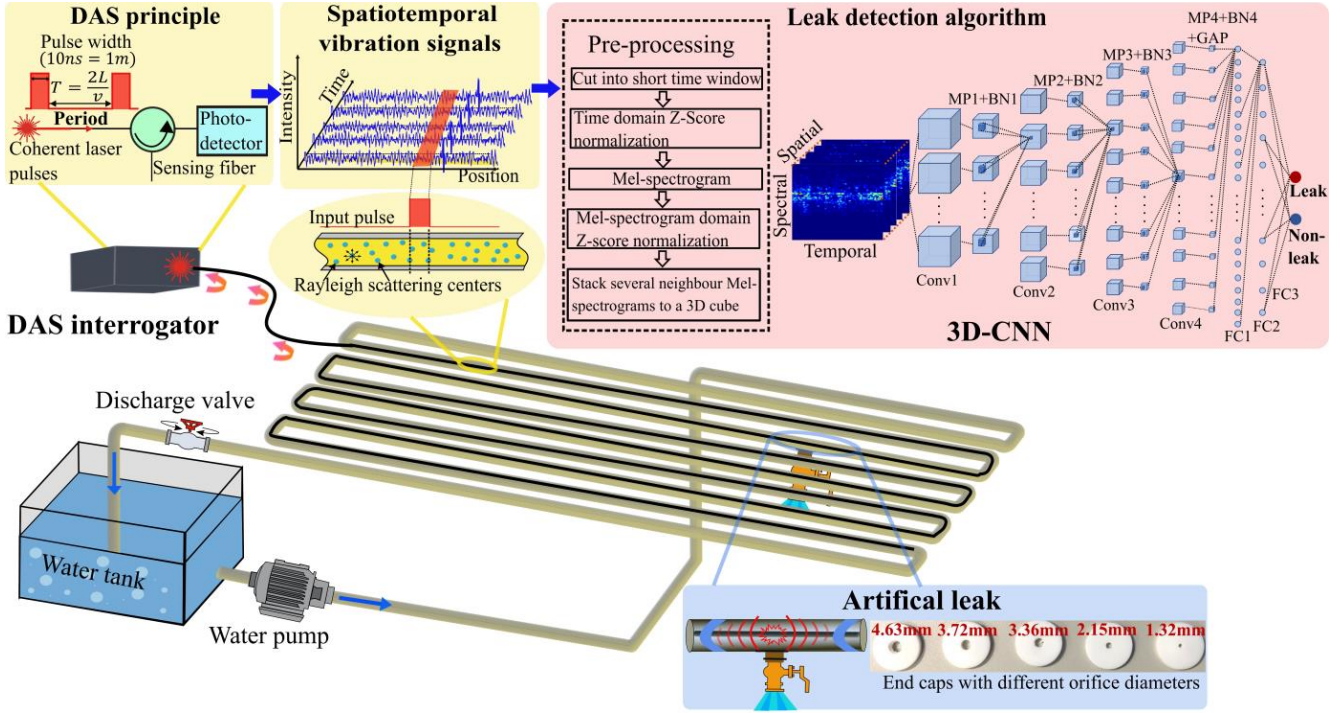

Figure 2. Testbed illustration of the proposed leak detection scheme.

## II. METHOD AND EXPERIMENT

The designed pilot experiment setup is illustrated in Figure. 2. The testbed includes a recirculating water pipe, a DAS system, and a leak detection algorithm.

### A. Recirculating water pipe testbed

The experiment was conducted at The Hong Kong Polytechnic University's Hydraulics Laboratory using a 60 m pipe with a 50 mm internal diameter and a 3.6 mm thick galvanized iron wall. Water was supplied from a reservoir by a 3-phase induction motor pump set at 2890 rpm. The pipe flow was varied between 0.4 L/s and 1.8 L/s. Five pressure gauges were installed on the pipe to measure the pressure and an ultrasonic flowmeter was installed downstream of the pipe. A GYFTY fiber-optic cable with non-metallic strengthening member, grease-filled polyethylene sheathed outdoor layers was attached on the outside pipe surface with nylon stripe and duct tape with 1 m interval. The cable, with an outer diameter of 8 mm and weight of 85 kg/km, contains four $9/125\ \mu m$ single-mode optical fibers, with one core used for sensing. The cable was not mounted on the beginning and end sections of the water pipe to avoid water inlet and outlet interference. The sensing length of the cable was 40 m. To simulate different leak states, an artificial leak was installed at the front section of the pipe, and five end caps with drilled orifice diameters of 4.63 mm, 3.72 mm, 3.36 mm, 2.15 mm, and 1.23 mm were used to simulate different leak levels. The relationship between the pipe position and the DAS channel number was determined by tap-tests.

### B. Principle of using fiber-optic cable for water pipe leak detection

Two types of vibrations can occur along a pipe: internal flow-induced pipe vibrations and leak-induced vibrations. Internal flow-induced pipe vibrations are caused by the interaction between water molecules and the pipe wall, and their strength is proportional to the square of the flow rate [9]. On the other hand, leak-induced vibrations are created when pressurized water pipes develop leaks, and high-speed water jets shoot out through the leak orifice, causing friction and cavitation-induced vibrations. The detection of these leaks is accomplished by identifying the signature of the continuous leak-induced disturbance to the surroundings. Fiber-optic cable can sense these vibrations is because phase change of Rayleigh backscattering has a linear relationship with the external vibration induced axial strain change. The localization is based on time-of-flight measurement principle in DAS. The system used in this work is an in-house built heterodyne detection DAS based on phase-sensitive OTDR technology as described in [10].

### C. Dataset collection

The proposed system's performance was evaluated using a dataset, described in Table II (a). The water pipe system was first measured with two flow rates (0.427 L/s and 1.80 L/s) for 28 minutes each without any leaks. Next, steady-state leaks were acquired for 14 minutes per case with the artificial leak valve turned on and different end caps in place. The leak flow velocity was measured by the volumetric method. The system defined three levels of leaks based on the ratio of leak flow rate to pipe flow rate: excessive leak (over 15%), significant leak (5-15%), and small leak (below 5%). The Reynolds number (Re.)

Table II (a). Dataset description

| Leak level | Case No. | Pipe flow rate(L/s) | Leak flow rate (L/s) | Leak ratio (%) | Orifice dia. (mm) | Pipe flow Re. | Leak flow Re. | Re. ratio | Duration (min) |
|---|---|---|---|---|---|---|---|---|---|
| Excessive leak | 1 | 0.427 | 0.319 | 74.7 | 3.72 | 10837 | 108987 | 10.06 | 14 |
| | 2 | 0.427 | 0.261 | 61.1 | 3.36 | 10837 | 98657 | 9.10 | 14 |
| | 3 | 0.427 | 0.102 | 23.9 | 2.15 | 10837 | 60418 | 5.58 | 14 |
| | 4 | 1.800 | 0.399 | 22.2 | 4.63 | 45674 | 109347 | 2.39 | 14 |
| Significant leak | 5 | 1.800 | 0.257 | 14.3 | 3.72 | 45674 | 87663 | 1.92 | 14 |
| | 6 | 1.800 | 0.209 | 11.6 | 3.36 | 45674 | 78918 | 1.73 | 14 |
| | 7 | 0.427 | 0.034 | 8.0 | 1.22 | 10837 | 35233 | 3.25 | 14 |
| Small leak | 8 | 1.800 | 0.084 | 4.7 | 2.15 | 45674 | 49696 | 1.09 | 14 |
| | 9 | 1.800 | 0.027 | 1.5 | 1.22 | 45674 | 28445 | 0.62 | 14 |
| No leak | 10 | 0.427 | - | - | - | 10837 | - | - | 28 |
| | 11 | 1.800 | - | - | - | 45674 | - | - | 28 |

Table II (b). Leak identification performances of 2D- and 3D-CNNs

| Leak level | Case No. | | 2D-CNN | | | | 3D-CNN | | | | Location error (m) |
|---|---|---|---|---|---|---|---|---|---|---|---|
| | | | Z=3 | Z=5 | Z=7 | Z=9 | Z=3 | Z=5 | Z=7 | Z=9 | Z=5 |
| Excessive leak | 1 | TPR (%) | 99.7 | **100.0** | 100.0 | 100.0 | **100.0** | 100.0 | 100.0 | 100.0 | 2.80 |
| | 2 | | 97.3 | 98.2 | 99.4 | 99.1 | **100.0** | 100.0 | 100.0 | 100.0 | 1.60 |
| | 3 | | 97.9 | 97.0 | 97.9 | 99.0 | 97.3 | 98.8 | 98.2 | **99.0** | 0.80 |
| Significant leak | 4 | | 93.4 | 97.3 | 96.4 | 96.7 | 94.6 | **98.8** | 94.9 | 97.0 | 0.76 |
| | 5 | | 87.3 | 92.7 | 92.5 | 91.3 | 89.2 | **92.8** | 92.8 | 90.4 | 0.40 |
| | 6 | | 83.1 | 83.7 | 79.8 | 82.8 | 87.3 | **87.7** | 84.9 | 84.9 | 0.12 |
| Small leak | 7 | | 90.1 | 88.3 | 84.6 | 80.1 | 90.7 | **92.2** | 90.6 | 92.0 | 0.44 |
| | 8 | | 75.9 | 78.3 | 81.0 | 83.7 | 82.8 | **87.7** | 85.2 | 82.2 | 0.28 |
| | 9 | | 47.9 | 57.5 | 51.5 | 56.0 | 51.8 | **60.8** | 58.7 | 60.5 | 0.32 |
| No leak | 10 | FAR (%) | 0.27 | 0.25 | 0.18 | 0.19 | 0.27 | 0.24 | 0.21 | **0.17** | - |
| | 11 | | 2.15 | 2.73 | 3.20 | 2.35 | 1.71 | **1.70** | 1.78 | 1.74 | - |

* The localization error is measured by Z=5 with 3D-CNN

was calculated to describe the flow status in the pipes. This dimensionless quantity provides insight into the relative importance of the inertial forces to the viscous forces for a given flow condition [11]. Laminar flow occurs when Re. is smaller than 2000 and the water travels in a direction parallel to the pipe axis. Turbulent eddies in all directions are imposed on the axial flow when Re. is larger than 3500, causing significant additional vibration. Detecting and locating small leaks despite large flow rates is a challenging task in leak detection systems. Therefore, we conducted the experiments under harsh conditions where turbulent pipe flows were fully developed for all cases. The total cable length was 100 m, with 40 m mounted along the water pipe and 60 m used for connection. The spatial resolution was set to 2 m and channel spacing was 0.8 m and the sampling frequency was 10 kHz in DAS. The data were collected on 20, and 21 Nov 2021. The raw data comprise over 100 GB with a total duration of 182 minutes.

*D. Leak detection algorithm*

In recent years, deep learning has outperformed traditional signal processing methods in a variety of fields. In the WDN domain, deep learning has been also explored for tasks such as demand forecasting, leak detection and localization, and water quality anomaly detection [12]. Convolutional neural networks (CNNs) have emerged as the most widely used deep learning method for leak detection and localization, using either flow/pressure data or acoustic/vibration data [12]. Since our data have spectral, temporal, and spatial dimensions due to the distributed sensing mechanism, we propose using 3D-CNNs, which jointly utilize these three dimensions for signal processing. To verify the effectiveness of leveraging inter-spatial channels by 3D-CNNs, we also trained 2D-CNNs with similar network architecture and input data for comparison.

**Pre-processing:** Traditionally, leak detection involved experienced operators who used mechanical listening sticks to discern leak noises. To replicate this method using automatic algorithms, we utilize a human auditory system that mimics Mel-spectrogram as the input feature vectors. The pre-processing steps, depicted in Figure 2, are as follows: (1) continuous DAS time series data are divided into 5-second segments for each space channel, (2) the 5-second signal clip is transformed to Mel-spectrograms with 128 bands covering the frequency range (0-5 kHz) after Z-score normalization, and the first 90 bands are used. (3) a window size of 204.8 ms and a hop length of 51.2 ms are applied, (4) Z-score feature scaling method is performed on each Mel-spectrogram to standardize the features, and (5) Mel-spectrograms from several neighboring positions are stacked to form a 3D cube.

**Network architecture:** The networks are designed to predict the leak probability of each 3D Mel-spectrogram cube. The input dimensions are $90 \times 98 \times Z$, where Z is the number of spatial channels of the input cube and Z=3,5,7,9. The networks have four 3D convolutional (Conv) layers interleaved with four max-pooling layers and batch normalization (MP+BN) operations. The kernel size and filter number in each layer are given in Table III. The rectifier linear unit (ReLU) is utilized for the non-linear activation function. The outputs of FC3 are mapped to the two-class label, leak, and non-leak. The softmax function is used to provide a probabilistic output.

**Training:** In each case, 75% of the data in Table II (a) are used for training and 25% are used for testing. For non-leak training data, signals from 7 positions including flange joint (1 position), elbow (2 positions), and straight pipe (4 positions) are chosen.

For the leak training data, signals at artificial leak positions are used. During the training process, the model optimizes binary cross-entropy via Adam optimizer algorithm with exponential decay. A mini-batch size of 128 is set and L2 norm regularizers with a penalty of 0.003 are applied after each convolutional layer to reduce overfitting. The model is trained with 100 epochs with early stopping criteria. The networks are implemented in Python using Tensorflow v2.4.1 libarary.

**Testing:** The test data include both non-leak and leak signals from all positions and the predictions for each 3D cube are used to create leak probability maps. The threshold for determining a leak is set at 0.9.

**Table III. Architecture of 2D-CNNs and 3D-CNNs**

| Layer | 2D-CNN $Z=3, 5, 7, 9$ | | 3D-CNN $Z= 3, 5, 7, 9$ | |
|---|---|---|---|---|
| | Kernel | Channel | Kernel | Channel |
| Conv1 | $3 \times 3$ | 16 | $3 \times 3 \times 3$ | 16 |
| MP1+BN1 | $2 \times 2$ | 1 | $2 \times 2 \times 1$ | 1 |
| Conv2 | $3 \times 3$ | 32 | $3 \times 3 \times 3$ | 32 |
| MP2+BN2 | $2 \times 2$ | 1 | $2 \times 2 \times S$ | 1 |
| Conv3 | $3 \times 3$ | 64 | $3 \times 3 \times 3$ | 64 |
| MP3+BN3 | $2 \times 2$ | 1 | $2 \times 2 \times 1$ | 1 |
| Conv4 | $3 \times 3$ | 128 | $3 \times 3 \times 3$ | 128 |
| MP4+BN4 | $2 \times 2$ | 1 | $2 \times 2 \times 2$ | 1 |
| Dropout | - | - | - | - |
| GAP | - | - | - | - |
| FC1 | $12 \times 128$ | - | $128 \times 128$ | - |
| FC2 | $128 \times 64$ | - | $128 \times 64$ | - |
| FC3 | $128 \times 2$ | - | $128 \times 2$ | - |

\* $S = 1$ for $Z = 3$, $S = 2$ for $Z = 5,7,9$.

## III. RESULTS AND DISCUSSIONS

### A. Leak identification and localization

To evaluate the leak identification performance, we used true positive rate (TPR) and false alarm rate (FAR). TPR is the ratio of correctly identified labeled leaks over all true leaks, while FAR corresponds to the ratio of non-leaks incorrectly identified as leaks over all true non-leaks. The leak identification results are summarized in Table II (b). With the same spatial range $Z$, 3D-CNNs outperform 2D-CNNs in most cases for both FAR and TPR, demonstrating that the inter-channel features could provide useful neighboring information and thus improve the identification performance [13]. The additional spatial information along the fiber contributes to the results in the form of a majority vote, making the system less vulnerable to environmental noises. Therefore, the 3D-CNN is more suitable to process the time-position correlated signal. Among the 3D-CNNs, $Z = 5$, which corresponds to a 4 m spatial range, showed the best overall performance. The TPR is found to range from 60.8% to 100% and is mainly determined by the leak flow over pipe flow ratio. High TPRs normally could have been achieved at high ratios since the leak-induced vibrations dominated around the leak position. Low TPR is observed for small leak-induced vibrations due to the background pipe flow noise. The FARs are 0.24% and 1.70% for pipe flow of 0.427L/s and 1.800 L/s, respectively. It is intuitive that high pipe flow causes more severe background noise and resulted in more false alarms.

After identifying the leaks, the location of the leak should be pinpointed for remedial action. Unlike traditional sparse sensor systems that requires specific features, leak localization can be easily determined from a leak probability in the proposed scheme. The leak probability map along the fiber and the time, predicted by the 3D-CNN with $Z = 5$, under six conditions are shown in Figure. 3(a). The median value of the leak probability along the 40 m cable over 210 seconds is plotted on the top of each map. Some false alarms are scattered around the map in case 10 and 11. The FARs can however be eliminated by considering data over a longer time, as shown in median probability plots (on top of each probability map). External perturbations may introduce noises but most of them do not last long, whereas the leak-induced vibrations continuously generate signals. This characteristic greatly helps distinguish leaks from environmental noises. As shown in Figure. 3(a) case 1, case 3, case 6, case 9, though some false negative predictions cause discontinuity, especially in significant leaks and small leaks, the leak position could still be located after calculating the long-term median probability. The leak flow rate as low as 0.027 L/s is successfully located. The location error is quantified using a threshold of 0.9 to determine the spatial range of the leak, and the middle position is considered the leak center. The location error is summarized in the last column of Table II(b). The error is between 0.12 to 2.8 m. Under excessive leak conditions, the leak-induced vibrations affect a wide range of the pipe, causing higher uncertainty in leak localization. While under significant and small leak conditions, the affected spatial range is relatively limited.

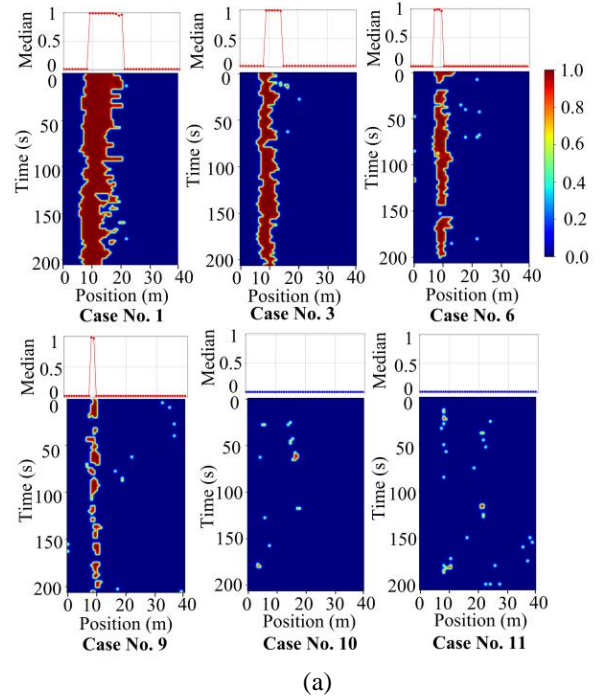

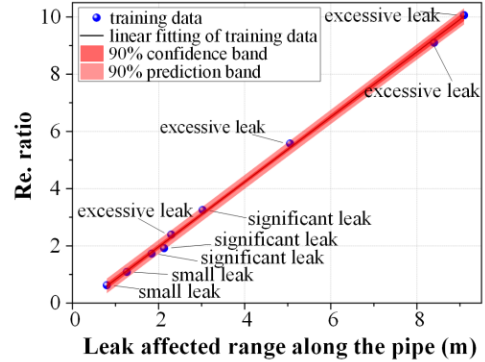

Figure 3. (a) Leak probability map of six cases, (b) Linear correlation of Re. ratio of leak flow and pipe flow (c) Truth table of leak level quantification.

*B. Leak level quantification*

Leak level quantification is equally important as leak identification and localization because it provides crucial information for repair prioritization. The extent of a leak's impact on a pipe is influenced by its flow rate relative to the pipe's flow rate, as shown in Figure 3(a). A correlation between the leak affected range and the leak flow Re. over pipe flow Re. is established using linear regression on the training data, as plotted in Figure 3(b). The fit produced an $R^2$ value of 0.999, confirming a strong correlation between the leak-affected range and Re. ratio. To calculate the leak flow rate, we introduced the orifice equation, which describes the conversion of pressure energy to kinetic energy in a water pipe leak [14]. Therefore, by combing the leak-affected range and Re. ratio relation and orifice equation, we can predict both leak diameter and leak flow rate. During the prediction stage, we calculated the mean leak affected range with a duration of 30 seconds for a given test dataset. As shown in Figure 3(c), we evaluated the performance of the leak quantification method by classifying the predicted leak levels using a three-level scale, achieving a classification accuracy of over 85%.

*C. Discussions*

**Optical network and WDN integration**: The transmission range for Super-PON can reach 50 km, which is comparable with the sensing length of DAS. The PON and WDN architectures are typically implemented in a point-to-multipoint topology, and to sense multipath simultaneously with one interrogator in a cost-effective way, multipath DAS based on frequency division multiplexing can be adopted [15]. Further research with large-scale experiments will be conducted at Q-Leak underground water mains leak detection training center and Anderson Road Quarry Site in Hong Kong to explore the challenges and opportunities of integrating optical networks and WDNs.

**Fiber-optic cable deployment and repair**: Deploying and repairing fiber-optic cables is crucial in the proposed scheme as the fiber itself is the sensor and the deployment method can significantly impact its performance and durability. Further research should focus on developing innovative installation techniques and materials that can withstand harsh environments and reduce the damage from external factors. Repair issues of fiber-optic cable should also be considered in future research. While integrating optical networks and WDNs can reduce the cost of civil works, it also increases the risk of physical damage to the cable. Although cables are usually designed to withstand some physical stress, pipe bursts and water hammer are potential risks that could cause damage. To mitigate these risks, protective casing or avoiding high-risk areas could be considered.

**Leak detection algorithm**: In real-world WDNs, vibrations induced by leaks can be much more complex and dynamic, with varying pipe materials, sizes, as well as different flow rates and pressures. Supervised learning with large datasets and powerful computer resources, has proven effective in classifying leak and non-leak cases with labelled data. However, to fully realize the potential of deep learning in leak detection, several challenges related to data and algorithmic development need to be addressed. The high cost of manual data labelling is a significant obstacle to developing accurate and reliable deep learning models. Semi-supervised learning with a small amount of labelled data and unsupervised learning with unlabelled data may be solutions to reduce labelling cost. Another challenge is the lack of open datasets and shared models, which can hamper the adoption of this technology in the water industry. Close collaboration between researchers and the water industry can help overcome these challenges and facilitate the adoption of deep learning for leak detection in WDNs. In addition, the use of hybrid techniques, combining different sensing modalities could further enhance the accuracy and reliability of leak detection and localization in WDNs.

## IV. CONCLUSIONS

The proposed approach has demonstrated the potential of utilizing data from an optical communication cable installed along a water pipe and using 3D-CNN for automatic leak detection. In the experiment, a 40-meter-long water pipe and an optical communication cable with a DAS interrogator were used to validate the technique's efficacy. Our proposed method successfully detected a leak flow rate as low as 0.027 L/s with a location accuracy of 3 meters and quantification accuracy of over 85%. The utilization of optical fiber sensing for water pipe leak detection has significant practical applications in the water industry, providing a new solution to a pressing issue of reducing water loss from aging pipes and declining freshwater resources. The ability to identify, locate, and quantify leaks can lead to substantial cost savings for water companies and improve overall water sustainability. Furthermore, the integration of optical fiber sensing and deep learning algorithms also opens new avenues for innovation in the field of leak detection in water industry.

## ACKNOWLEDGEMENTS

This research was funded by the Research Grants Council of Hong Kong under project numbers 15209919, 152164/18E, 152007/19E, 152233/19E, 15200719 and The Hong Kong Polytechnic University (YW3G, ZVGB, BBWB, postdoc matching fund). The authors are grateful to Mr. Victor Lo for his valuable insights on leak detection and to Mr. Kwok Hing

Leung's for his support in conducting the experiments. Additionally, the authors acknowledge the PolyU University research facility in big data analytics for providing the computing resources.

BIOGRAPHIES

**Huan Wu** received B. Eng degree from the Nanjing University of Aeronautics and Astronautics in 2013 and Ph.D. degree from the Chinese University of Hong Kong in 2018. She is a postdoctoral researcher with the Department of Land Surveying and Geo-Informatics at The Hong Kong Polytechnic University. Her research interests include optical fiber sensing and its applications.

**Huan-Feng Duan** received Ph.D. degree in Hydraulics from the Hong Kong University of Science and Technology in 2011. He is currently an Associate Professor in the Department of Civil and Environmental Engineering at The Hong Kong Polytechnic University. His research interests include pipe defects detection, urban hydraulics, coastal disaster analysis, and ocean renewable energy. He was the recipient of the 2022 Karl Emil Hilgard Hydraulic Prize from the American Society of Civil Engineers (ASCE).

**Wallace W. L. Lai** is currently an Associate Professor and Associate Head (Teaching) with the Department of Land Surveying and Geo-Informatics, The Hong Kong Polytechnic University. His main research interests include survey and mapping, 3-D imaging, and diagnosis of engineering structures, underground utilities and especially leakage, and construction materials.

**Kun Zhu** is a research fellow at The Hong Kong Polytechnic University. He received his B.Eng. and Ph.D. degrees both from Zhejiang University in 2007 and 2012. His research interests include distributed fiber sensing systems, signal processing based on optical fiber and waveguide devices, and microwave photonics.

**Xin Cheng** received his M.S. and Ph.D. degrees from the Northeast Normal University and Jinan University in 2009 and 2019. He is a Post-doc fellow in The Hong Kong Polytechnic University Photonic Research Center (PRC). His research interests are specialty optical fiber design and its applications.

**Hao Yin** received M.S. degree from the City University of Hong Kong in 2021. He is a Research Assistant in The Hong Kong Polytechnic University.

**Bin Zhou** received his B.Sc. and Ph.D. degree from Physics Department of Zhejiang University in 2006 and 2011. He is an Associate Professor at South China Normal University. His research interests are in the areas of optical sensing, optical devices and nonlinear optics.

**Chun Cheung Lai** obtained his PhD from The Hong Kong Polytechnic University. Currently, he is a Scientific Officer in the Department of Civil Engineering. His research interests include sensor design and robotics.

**Chao Lu** obtained his BEng in Electronic Engineering from Tsinghua University, China in 1985, and his MSc and PhD from University of Manchester in 1987 and 1990 respectively. He is currently Chair Professor of Fiber Optics in the Department of Electronic and Information Engineering, The Hong Kong Polytechnic University. His current research interests are in the area of high-capacity transmission techniques for long haul and short reach systems and distributed optical sensing systems.

**Xiaoli Ding** obtained BEng in Surveying from Central South University, China in 1983 and PhD from University of Sydney, Australia in 1992. He is Chair Professor of Geomatics in the Department of Land Surveying and Geo-Informatics, The Hong Kong Polytechnic University. His main research interests include radar remote sensing, GNSS, geohazards and land development.